\documentclass[pre,amsmath,twocolumn,amssymb]{revtex4}
\usepackage{epsfig}
\usepackage{graphicx}
\usepackage{dcolumn}
\usepackage{bm}

\begin{document}

\title{Flow boundary conditions from nano- to micro-scales}

\author{Lyd\'eric Bocquet}
\affiliation{Universit\'e Lyon I Laboratoire de Physique de la
Mati\`ere Condens\'ee et des Nanostructures;  CNRS, UMR 5586, 43
Bvd. du 11 Nov. 1918, 69622 Villeurbanne Cedex, France}
\author{Jean-Louis Barrat}
\affiliation{Universit\'e de Lyon; Univ. Lyon I,  Laboratoire de
Physique de la Mati\`ere Condens\'ee et des Nanostructures; CNRS,
UMR 5586, 43 Bvd. du 11 Nov. 1918, 69622 Villeurbanne Cedex,
France}
\date{\today}
\begin{abstract}
The development of microfluidic devices has recently revived the
interest in "old" problems associated with transport at, or
across, interfaces. As the characteristic sizes are decreased, the
use of pressure gradients to transport fluids becomes problematic,
and new, interface driven, methods must be considered. This has
lead to new investigations of flow near interfaces, and to the
conception of interfaces engineered at various scales to reduce
flow friction. In this review, we discuss the present theoretical
understanding of flow past solid interfaces at different length
scales. We also briefly discuss the corresponding phenomenon of
heat transport, and the influence of surface slip on interface
driven  (e.g. electro-osmotic) flows.

\end{abstract}
\maketitle

\section{Introduction}

The nature of boundary conditions of fluid at solid surfaces has
been revisited over the recent years. Beyond the fondamental
understanding of the fluid-solid dynamics, the reason for such a
strong interest lies in its potential applications in
microfluidics. The driving of liquids in ever tiny channels raises
a number of difficulties, one being the huge increase in
hydrodynamic resistance when the channel size decreases. Releasing
the no-slip boundary condition at the surfaces, and thereby
allowing for boundary slippage, would allow to bypass this
stringent conditions by decreasing wall friction. There is
therefore a big hope to take benefit of slippage for microfluidic
applications.

Many experiments have been performed on the subject, with
sometimes contradicting results. We refer to the review of Lauga,
Brenner and Stone for an exhaustive discussion of the experimental
approaches to investigate slippage \cite{lauga05}. While
experimental investigations  focused first on slippage on bare
(atomically smooth) surfaces, more  recent works have turned
towards structured  surfaces, in particular superhydrophobic
surfaces, characterized by patterns at the micro- or nano-scales.

In this review we concentrate on the theoretical understanding and
expectations for the fluid-solid boundary condition. In
particular, we will discuss the relevant theoretical framework for
  mechanisms that take place at different scales, and  lead to slippage at
  solid-fluid interfaces.

\section{Flow past ideal interfaces/}
\label{ideal}

From a conceptual standpoint, the simplest situation that can be
considered is that of a semi infinite solid, bounded by an
atomically smooth surface, in contact with a liquid that occupies
the second half space. By atomically smooth, we mean for example a
dense plane of a perfect crystalline lattice. The location of this
lattice plane defines the $xOy$ plane. The theoretical question
becomes then simply to describe momentum transfer from the fluid
to the solid (assumed here to be at rest) in terms of the velocity
field existing in the fluid. Let us consider the case, where the
fluid is undergoing a laminar, planar Couette flow  in the $x$
direction, e.g. driven by a second flat boundary far away in the
$z$ direction. In a stationary state, the stress component
$\sigma_{xz}$ is uniform, and the velocity profile of a Newtonian
fluid reads, far away from the solid wall
\begin{equation}\label{profile}
    v_x(z) = \dot{\gamma} z + v_s
\end{equation}
The shear stress  is $\eta \dot{\gamma}$, where $\eta$ is the
shear viscosity of the bulk fluid and $\dot{\gamma}$ the shear
rate. By definition, $v_s$ is the slip velocity. To achieve a
complete hydrodynamic description of the system, a constitutive
equation involving the interface has to be introduced. Such a constitutive equation is
generally described as a "boundary condition". Although we will
also use this traditional vocabulary, we emphasize that it is
somewhat misleading. In many cases,  "boundary conditions" are
seen  as auxiliary mathematical constraints that allow to solve
partial differential equations.  In our mind, the status of the
constitutive equation relative to the interface is in every
respect comparable to the one of the bulk equations, e.g.

\begin{itemize}
\item it is a material property, involving  the chemical nature of
the two phases that create the ideal interface.

\item it should be amenable to the same type of microscopic
justifications and statistical mechanical treatments as bulk
properties.

\item it can be extracted unambiguously from transport
measurements.

\end{itemize}

\begin{figure}[t]
\centerline{\resizebox{7cm}{!}{\includegraphics{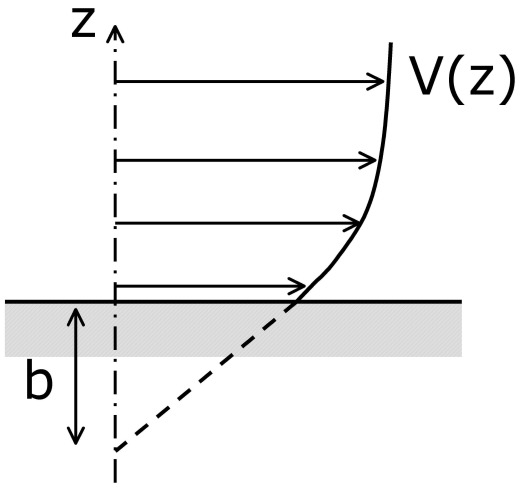}}}
\vspace*{-0.2cm} \caption[]{Schematic representation of the
definition of the slip length $b$} \label{fig1}
\end{figure}

As for any constitutive equation, the writing of the interfacial
constitutive equation involves a first step of modeling based on
considerations of symmetry and linearity. The simplest possible
relation, consistent with the general description for newtonian
fluids, assumes that the tangential stress exerted on the solid
surface is proportional to the slip velocity, i.e.
\begin{equation}\label{bc1}
   \sigma_{xz}  = \kappa  v_s
\end{equation}
Combining (\ref{bc1}) with the constitutive equation for the bulk
Newtonian fluid, $\sigma_{xz}= \eta \partial_z v_x$, one arrives
at the so called Navier boundary condition
\begin{equation}\label{bc2}
     v_s = \frac{\eta }{\kappa} \partial_z v_x = b \partial_z v_x
\end{equation}
which can be used as a boundary condition (in the mathematical
sense) that complements the Navier-Stokes equation inside the
fluid. This boundary condition is applied within one atomic
distance  of the first solid plane \cite{bocquet94}, where the
stress (\ref{bc2}) acts on the solid. By construction, the
"hydrodynamic" flow that results from using this boundary
condition will be identical to the actual, macroscopic flow far
away from the boundaries. At the microscopic level, a thin
interfacial region exists in which the flow may differ from the
hydrodynamic calculation. We insist that this region, in which the
velocity field can be obtained only from molecular scale studies,
is not our main concern here. Studies at the molecular scales may
result into velocity profiles that are not solution of the Stokes
equation, or need to artificially introduce a position dependent
viscosity.

The second equality in  (\ref{bc2}) introduces the {\em slip length}
$b=\eta/\kappa$. The significance of this length is illustrated in
figure \ref{fig1}, as the distance inside the solid to which the
velocity profile has to be extrapolated to reach zero. The
standard "no slip" boundary condition corresponds to $b=0$, i.e.
$\kappa\rightarrow \infty$. Conventionally a positive slip length
is associated with a positive slip velocity, while a (less common)
negative length would indicate an apparent change in the sign of
the velocity field near the solid. We emphasize that $b$ is not an
interfacial property in the usual sense, rather it is the ratio of
a truly interfacial property, the friction $\kappa$, to a bulk
property, the viscosity $\eta$.  In particular, $b$ is, in
principle, completely unrelated to the width of the interfacial
region alluded to above. All our knowledge of interfaces between
solids and simple liquids indicates that -if critical phenomena
are excluded- the width of the interfacial region is at most of a
few atomic diameters. The slip length, on the other hand, can
(theoretically) be made to diverge by going to the limit of a
perfectly flat (in the mathematical sense) solid wall. Such a
procedure, on the other hand, would not significantly alter the
width of the interfacial region.

The determination of the interfacial constitutive relation, often
described as "measurement of the slip length", has been the
subject of numerous experimental
\cite{lauga05,Chan85,georges93,pit00,cecile04,cecile05} and
theoretical
\cite{thompson90,bocquet94,barrat99a,cieplak01,zhu01,priezjev04,jabbarzadeh06}
investigations in the past 20 years . In spite of the experimental
difficulties associated with the definition of "ideal" surfaces in
experiments, the combination of increasingly accurate measurement
methods and of insights from simulation and theory has produced a
rather complete understanding of the "ideal" situation discussed
here. The results can be summarized as follows.

(i) The {\em linear} relation (\ref{bc2}) can be used to describe
results for simple liquids, within the range of experimentally
accessible shear rates \cite{cecile05}. Numerically, non linear
behavior has been reported for shear rates of order $10^8 s^{-1}$ for
simple liquids
\cite{thompson97}.

(ii) The friction coefficient $\kappa$, and hence the slip length
$b$,  depend strongly on the strength of the solid liquid
interactions. Early measurements \cite{Chan85,georges93} were
performed on fluid/solid interfaces that correspond to a "wetting"
situation, with a low contact angle. These measurements clearly
reported a "no slip" boundary condition, with a slip length
smaller than the molecular dimensions.

(iii) When the interaction between liquid and solid becomes
weaker, so that the contact angle of the liquid on the substrates
increases, larger values of $b$ can be measured, typically in the
range 10-50 nanometers.

(iv) Points (ii) and (iii) correspond to a general trend, but
there is not a unique correlation between contact angle and
friction. Other factors, such as  the crystallinity of the solid,
do influence the friction $\kappa$ \cite{thompson90,thompson92}.

These findings can be rationalized within the context of a linear
response theory \cite{bocquet94,barrat99a} (see also
\cite{smith96}), which has been successfully compared to molecular
simulations \cite{barrat99a,priezjev04}. A possible starting point
is the Kubo like formula, which relates the interfacial transport
coefficient for momentum (i.e. the friction $\kappa$) to the
integral of the autocorrelation function of the momentum flux
(i.e. the force exerted by the liquid on the solid):
\begin{equation}\label{kubo}
\kappa =\frac{S}{k_BT} \int_0^\infty dt \langle f_{x}(t) f_{x}(0)
\rangle
\end{equation}
where $S$ is the surface area, and $f_{x}(t)$ is the tangential
stress (force per unit area) exerted by the fluid on the solid at
time $t$ in the $x$ direction\footnote{A matrix generalization of
\ref{kubo} is easily obtained for anisotropic surfaces.}. Note
that this formula can obviously be generalized to the case where
$\kappa$ would be anisotropic, represented by a $2\times 2$
matrix, as would be the case for a surface of low symmetry. We
will limit ourselves to the standard case of isotropic friction.
Kubo formula such as (\ref{kubo}) cannot in general be evaluated
analytically. The correlation function can be computed in direct
numerical simulations, or approximated using the standard methods
of liquid state theory.

A simple approximation of the result of equation  (\ref{kubo}) was
derived in \cite{barrat99a}, and allows one to quantify the main
ingredients that enter the friction. The approximate expression
reads
\begin{equation}\label{kubo-app}
    \kappa \backsimeq  \frac{S_\parallel(q_0)}{D_\parallel k_BT }
    \int_0^\infty dz \rho(z) V_{FS}(z)^2
\end{equation}
Here $q_0$ is the first wave-vector of the reciprocical lattice of
the crystalline substrate; $S_\parallel(q_0)$ is the structure
factor of the fluid evaluated for this wave-vector, in the
interfacial region; $D_\parallel$ is a collective diffusion
coefficient computed at the wave-vector $q_0$; $\rho(z)$ is the
density profile of the fluid perpendicular to the interface, and
$V_{FS}(z)$ is the interaction potential between a molecule in the
fluid and the solid wall. The physical content of the various
factors entering (\ref{kubo-app}) is clear. The structure factor
describes the response of the fluid to the fixed atomic
corrugation of the solid wall. A large $S_\parallel(q_0)$
indicates "commensurability" of some sort between the liquid and
the solid, which will increase momentum transfer. The diffusion
constant sets, roughly,  the time scale for the decay of
stress-stress correlations.

Note that when $\kappa$ is divided by $\eta$ to obtain $1/b$, the
product $D_\parallel\eta$ will be formed. For simple liquids, this
ratio is typically equal to $k_B T $ divided by a molecular size
$\sigma$, so that any explicit reference to the microscopic
dynamics disappears from the resulting expression for the slip
length.
\begin{equation}\label{b-app}
    b \backsimeq \frac{(k_BT)^2}{ S_\parallel(q_0) \sigma
    \int_0^\infty dz \rho(z) V_{FS}(z)^2}
\end{equation}
 It is therefore
expected to be independent of the fluid viscosity for simple
liquids.

The last important factor is  the integral, which involves the
wall fluid interaction and the density profile. It is through this
integral, that the wetting properties of the fluid with respect to
the solid enter the friction $\kappa$. As a crude approximation,
let us assume that the density profile consists of a layer of
density $\rho_c$ (the "contact density") within a molecular
thickness $\sigma$, followed by a bulk density $\rho_B$. We
moreover assume van der Waals interactions with Hamaker constant
$A_H$.  Then the integral can be crudely  approximated by
$(A_H^2/\rho_B\sigma^2) \times (\rho_c/\rho_B)$.

Gathering the different ingredients that enter equation
(\ref{kubo-app}), it is easily seen that the only one that can
display large variations is the "contact density", i.e. the
average density of the liquid very close  to the solid surface,
typically within one molecular size, and, to some extent, the
ratio $A_H/k_BT$. Both factors go into the same direction, namely
large slip lengths are favored by a small liquid solid
interaction. However, it is clear that, from a dimensional
viewpoint, the integral $\int_0^\infty dz \rho(z) V_{FS}(z)^2$
will always remain a "molecular" quantity, so that the length
scale that emerges from equation (\ref{b-app}) is not expected to
become very large unless very special state points are considered.
In fact, our simulations show that the slip lengths that are
achieved for simple Lennard-Jones models vary between a few
molecular diameters and 50 to 60 molecular diameters, depending on
the value of the interaction parameter that defines the solid
liquid attraction (Hamaker constant). This is in very reasonable
agreement, at a qualitative level, with the most recent
experimental results \cite{cecile05}.

The same framework may be applied to melts of short polymers. In
this case, the substrate wave-vector $q_0$ is much larger than the
inverse radius of gyration of the polymer, so that the collective
diffusion is measured at the monomer scale : $D_\parallel$ is
accordingly expected to be independent of the molecular weight
$N$. This predicts that the ratio $D_\parallel\eta$ scales like
$\eta$ and increases with the polymer length, similar the
predictions of de Gennes for longer chains \cite{degennes79}. This
prediction and the validity of Eq. (\ref{kubo-app}) has been
exhaustively verified for short polymers (up to 16 monomers) using
Molecular Dynamics simulations by Priezjev and Troian
\cite{priezjev04}. The experimental situation may be more complex
since slippage will be strongly dependent on the polymer
adsorption at the surface. Polymers also display strong nonlinear
effects at moderate shear rates, which are not accounted for in
our description, and have been extensively studied experimentally
\cite{leger97,fetzer06} and theoretically \cite{brochardwyart94}.

Another interesting example of the use of equation (\ref{kubo}) is
the recent study of the influence of electrostatic effects
 on the solid-liquid friction in ionic solutions \cite{joly06,joly06-1}. A
simple estimate of these effects based on  Eq. (\ref{kubo}) leads
to an electrostatic friction contribution to the friction factor,
varying as $\kappa' \propto \eta \Sigma^2 \ell_B$, with $\Sigma$
the surface charge (in units of the elementary charge $e$), and
$\ell_B=e^2/4\pi\epsilon k_BT$ the Bjerrum length (with $\epsilon$
the dielectric constant). Interestingly this contribution is
independent of the Debye length.

 While the theoretical and numerical descriptions converge
unambiguously, a number of experimental reports have appeared over
the years, that would appear to contradict either (i) or (ii)
above, or both \cite{neto05,pit00,zhu01,bonaccurso03}. In some
cases, these discrepancies can be attributed to experimental
artefacts associated with the measurement apparatus
\cite{vinogradova03,cecile05}. However, another possible  source
of misunderstandings and errors lies in the structuration effects
described below. The {\em intrinsic} constitutive boundary
condition discussed in this section may be rather different from
what is probed in flow experiments at larger length scales.

We conclude with a brief discussion of a view of the solid liquid
interface which is sometimes proposed, in which the interfacial
region is described as a "film" of thickness  $w_i$, formed by
some interfacial fluid  (or vapor) of viscosity $\eta_i$. Standard
"no slip" boundary conditions are applied at the film boundary. A
straightforward calculation shows that, within this description,
the friction $\kappa$ that enters the constitutive equation for
the interface is given by
\begin{equation}\label{kappaeff}
    \frac{1}{\kappa} =  \left( \frac{1}{\eta_i}
    -\frac{1}{\eta}\right) w_i
\end{equation}
and the equivalent slip length is
\begin{equation}\label{beff}
    b  = w_i \left(\frac{\eta}{\eta_i}-1\right)
\end{equation}
Such a description does not have a clear microscopic funding. In
general, it  is  impossible to identify a "phase" with distinct
properties at the interface. However, it gives a useful insight
into the sensitivity of interfacial transport parameters to the
structure of the interface. For example, assuming that the
viscosity $\eta_i$ is that of a vapor layer, it is easily shown
that a slip length of 20nm corresponds to an interface thickness
of just  $0.25$ nm (for water), hardly the size of a single
molecule. Clearly this "layer" cannot correspond to a real phase,
but is merely a very schematic representation of the depletion
close to hydrophobic solid walls.

\section{Climbing the structuration scale}

\subsection{Effective slippage on patterned surfaces}
The situation of an atomically smooth surface, as described in the
previous paragraph, involves basically two lengths scales : the
microscopic scale, $\ell$, characterizing the liquid-solid
interface, typically of the order of the atomic size; and the
macroscopic size ${\cal L}$, corresponding to the flow scale, e.g.
the size of the channel in which the flow is conducted.

Intermediate scales may however be present, for example when the
surface exhibits large scale structuration, associated with
roughness or chemical patterns on a scale $L$. Assuming a scale
separation between the different lengths, $\ell \ll L <{\cal L}$,
the microscopic information associated with the liquid-solid
dynamics may be integrated out and summarized in the local slip
length on the patterned surface. In this case, the hydrodynamic
quantities, velocity profile and pressure, do obey the Stokes
equation, complemented by the local Navier condition with local
slip lengths $\{b\}$ (of order $\ell$) at the interface.

 Once the solution of this
set of equations has been calculated, it is possible to define an
effective boundary condition, by integrating out the intermediate
 scale $L$. This introduces an effective slip length $b_{\rm eff}$,
which may be defined in terms of the friction force $\mathbf{F}_f$
on the solid interface and an averaged slip velocity $ \langle
V_s\rangle$, as~:
\begin{equation}
\mathbf{F}_f= - {\cal A}\,  \eta \, { \langle V_s \rangle \over
b_{\rm eff}}
\end{equation}
The averaged slip velocity $ \langle V_s\rangle$ is computed in a
given plane parallel to the interface and the effective slip
length will thus depend on the specific choice for the location of
this plane  (e.g., top, bottom or mean of the roughness).

This definition of $b_{\rm eff}$ integrates out the
modulation of the surface at the scale $L$. The effective slip
length, $b_{\rm eff}$, thus defines the effective boundary
condition for the velocity profile $v$, {\it at scales larger than
$L$}~:
\begin{equation}\label{bceff}
     v_s = b_{\rm eff} \partial_z v_x
\end{equation}
The friction force on the surface can be obtained from the
pressure $P$ and viscous stress tensor
$\overline{\overline\sigma}$ at the solid interface ${\partial
\Sigma}$~:
\begin{equation}
\mathbf{F}_f= \int_{\partial \Sigma} \left( -P \mathbb{I}
+\overline{\overline\sigma} \right) \cdot d\mathbf{S}
\end{equation}
with $\mathbb{I}$ the unit tensor. From the previous definition,
the calculation of the effective slip length therefore reduces to
the resolution of the initial (pure) hydrodynamic problem,
involving the evaluation of the hydrodynamic velocity field and
pressure fields. These fields are obtained from Stokes equation,
complemented by the {\it local} boundary condition on the
interface, with local slip lengths $\{b\}$.

As a first example of such a procedure we consider the simple case
of   flow past a wavy solid surface. For simplicity we consider a
flow velocity perpendicular to the modulation of the surface, so
that the situation is effectively two dimensional. The effective
slip length is obtained in terms of the geometrical parameters of
the interface, the period $\lambda=2\pi/q$ and amplitude $u$ ($\ll
\lambda$) of the roughness :  $b_{\rm eff}={\lambda \over 2 \pi}
f(q u)$. In the case where a perfect slip boundary condition
applies locally on the surface, one obtains $f(q u) \propto(q u
)^{-2}$ \cite{degennes02}, while for a local no-slip boundary
condition on the surface, $f(q u ) \propto (qu)^2$
\cite{stroock02}. In both cases, the effective boundary condition
applies at the averaged position of the surface. Though simple,
these results provide some interesting insights into the nature of
effective boundary condition :
\begin{itemize}
\item  The result for the perfectly slipping interface shows that
roughness inevitably decrease slippage. The friction originates
accordingly in the viscous dissipation as the fluid flows past the
roughness pattern. This result was first pointed out by Richardson
\cite{richardson73}, who showed that the no-slip boundary
condition measured at the macroscopic scale follows from the
roughness of surfaces. Recent MD simulations confirm this result
\cite{cecile03,cecile04}.

\item On the other hand the result for a wavy surface with a local
no-slip boundary condition suggests an effective slip on a wavy
no-slip surface, while an increased dissipation is expected on a
modulated surface. This is a direct consequence of the choice for
the position of the boundary plane, where the effective boundary
condition applies : for the above results the averaged position
has been chosen. For a boundary plane position at the bottom of
the interface, the effective slip length is shifted by $-u$ and
thus negative, in agreement with the increased dissipation on the
modulated surface. This shows that the effective BC applies at
some position between the bottom and top of the roughness.

 \item
The above results applies for a single surface. If the flow is
conducted in a channel with two confining surfaces separated by a
distance $H$, the above expressions for the effective slip lengths
are modified, and do depend on $H$  ! (see eg \cite{stroock02}).
Such a dependence is exhibited when the gap size compares with the
roughness period $H\lesssim \lambda$. This points to the fact that
effective boundary condition is {\it not} a characteristic of the
liquid-solid interface solely, but may depend on the flow
configuration when the surface modulation scale $L\equiv \lambda$
becomes comparable (or larger) than the flow scale ${\cal L}\equiv
H$.
\end{itemize}


\subsection{Slippage on superhydrophobic surfaces}
Another situation which has been the object of a strong interest
over the recent years is that of superhydrophobic surfaces. These
surfaces combine   hydrophobicity at the molecular scale with
roughness at intermediate scales, leading to the so-called "Lotus
leaf effect" \cite{quere05}. The roughness amplifies the natural
non-wetting caracter of the surface, leading to very large contact
angles - up to 175$^\circ$ for a liquid drop on the surface. We
consider here the situation where the liquid surface remains at
the top of the roughness (Cassie state). The liquid surface is in
contact with the solid only through a small fraction $\phi_s$ of
the surface, while the remaining is free-standing. At the
hydrodynamic level, this composite surface leads to a spatially
dependent boundary condition with  a no-slip boundary condition on
the real solid-liquid interface (with fraction $\phi_s$), while
the remaining surface is characterized by a perfect slip boundary
condition ($b=\infty$) on the liquid-vapor interface.

Although the hydrophobicity of the surface at the molecular scale
tends (according to the discussion of section \ref{ideal}) to
increase the slip length at small scale, the essential role of
hydrophobicity is here a structural  rather than dynamical one.
The rough hydrophobic surface replaces part of the liquid solid
interface with a stress free liquid vapour interface.

\begin{figure}[t]
\centerline{\resizebox{7cm}{!}{\includegraphics{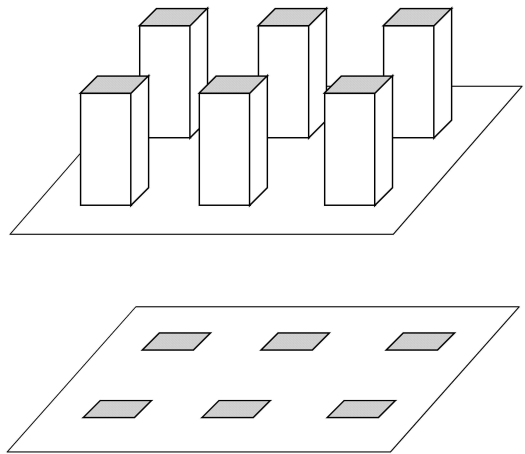}}}
\vspace*{-0.2cm} \caption[]{Schematic representation of a surface
patterned with nanoposts, and equivalent representation in terms of
flow boundary conditions.} \label{fig2}
\end{figure}

This problem has been first tackled theoretically by Philip
\cite{philip72} and more recently by Lauga and Stone
\cite{lauga03} and Cottin-Bizonne {\it et al.}
\cite{cottin-bizonne04}. Several numerical approaches have also
been proposed either at the molecular scale, using molecular
dynamics \cite{cottin-bizonne03}, or at larger mesoscopic scales
using finite element methods, Lattice Boltzmann or phase field
models \cite{priezjev05,benzi06,harting06}. For a pattern composed
of stripes, the expression for the effective slip length $b_{\rm
eff}$ depends on the direction of the flow. For a flow parallel to
the stripes, $b_{\rm eff}$ reads~:
\begin{equation}
b_\mathrm{eff}= \frac{L}{\pi} \log({1\over
\cos[\frac{\pi}{2}(1-\phi_s)]}) \label{blauga}
\end{equation}
For a flow in the direction perpendicular to the flow, the result
is given by the above expression divided by a factor $2$. These
expressions exhibit a weak (logarithmic) dependency on $\phi_s$.
In practice, the logarithmic term only varies in the interval
$[0.4;1.0]$ for $\phi_s$ between 0.1 and 5$\%$, {\it i.e.} for
feasible surfaces.

At a qualitative level, the theoretical results for the stripes
can be summarized by stating that the effective slip length
$b_\mathrm{eff}$ essentially saturates at the value fixed by the
lateral scale of the roughness $L$, with an unfavorable prefactor
($1/(2\pi$) in Eq. (\ref{blauga})). For example Eq. (\ref{blauga})
shows that a slip length larger than the  period of the pattern,
$b_\mathrm{eff}=L$, is  obtained  only when $\phi_s\leq 10^{-3}$.
A large slippage is therefore difficult to obtain within the
stripe geometry.

Another, more 'natural', geometry is that of a pattern of posts
(see figure \ref{fig2}). In this case the hydrodynamic model
consists in assuming a flat interface with a no-slip BC on the top
of the posts, while the remaining of the surfaces obeys a
shear-free BC.

No analytical resolution of Stokes equation with this set of
boundary conditions has been performed up to now. However a simple
'scaling' argument  accounts for the effective slip length in the
limit of a small fraction $\phi_s$ of solid zones \cite{ybert06}.
The friction force on the surface reduces to the force on the
solid zones $F_f= {\cal A} \phi_s \eta \dot \gamma_s$, with
$\dot\gamma_s$ the averaged shear rate on the solid patch. To
estimate $\dot \gamma_s$, one may note that the velocity profile
in the liquid is influenced by the solid zones only in a region of
their size, $a$, in all directions : this behavior is due to the
Laplacian character of the Stokes equation obeyed by the fluid
velocity. One therefore expects $\dot\gamma_s\sim U/a$, where $U$
is the slip velocity of the fluid on the free slip zones.
Therefore $F\simeq  {\cal A} \phi_s \eta U/a$. Now the effective
slip length is defined by  $F_f= {\cal A} V/b_{\rm eff}$, with
$V\approx U$ the slip velocity. One therefore deduces, in the
limit $\phi_s\rightarrow 0$,
\begin{equation}
b_{\rm eff}\propto {a\over \phi_s}.
\end{equation}
For a pattern made of individual posts, $\phi_s= (a/L)^2$, and
$b_{\rm eff}\sim L^2/a\sim L/\sqrt{\phi_s}$. A numerical
resolution of the Stokes equation with a pattern of no-slip square
dots confirms the validity of the heuristic argument above in the
limit of small $\phi_s$ (basically $\phi_s <40\%$) \cite{ybert06}.
It moreover gives the prefactor of the predicted relationship,
which is very close to $1/\pi$, so that \cite{ybert06}~ (in the
limit $\phi_s\rightarrow 0$):
\begin{equation}
b_{\rm eff}\simeq {1\over \pi} {L \over \sqrt{\phi_s}}
\label{slipplot}
\end{equation}
This simple prediction would deserve an analytical justification,
which has not been performed up to now.

We note that the above argument is independent of the geometry of
the pattern - posts or stripes -  and one should recover the
expression in Eq. (\ref{blauga})  for stripes. For solid zones
made of stripes, we have $\phi_s=a/L$, so that $b_{\rm eff}\sim L$
: up to slowly varying logarithmic terms, this is in good
agreement with the Philip-Lauga-Stone prediction, $b_{\rm eff}\sim
L/\log(1/\phi_s)$. As we emphasized above, in this geometry, the
effective slip length is mainly fixed by the roughness period,
$L$.

From the experimental point of view, several experimental studies
have been conducted to measure slippage effects on
superhydrophobic surfaces. Using pressure drop experiments,
Rothstein and coworkers  have reported slip lengths in the micron
range on patterns of stripes \cite{ou04,ou05}, with periodicity of
the stripe in the tens of microns range (up to $150\mu$m).
Experiments are in good agreement with the prediction of
hydrodynamic model of Philip \cite{philip72}. More recently Choi
{\it et al.} performed slip measurements on the same geometry but
with nanoscale patterns \cite{choi06},  showing a slip length of a
few hundred of nanometers, again in agreement with theoretical
hydrodynamic predictions.

On surfaces made of a pattern of pillars and using rheological
measurements, Choi and Kim \cite{choi06-b}  reported slip lengths
in the tens of micron range on a surface made of nanoposts,
$b_{\rm eff}\simeq 20\mu$m for water. The expression for the slip
length on dilute posts derived above, $b_{\rm eff}\simeq {1\over
\pi}ÊL/\sqrt{\phi_s}$, requires a very small solid fraction in
their case to reach  a $20 \mu$m slip length (with $L\sim 1\mu$m)
: $\phi_s\sim 0.03$\%. The contact angle on such surfaces with
ultra-low solid fractions is expected to be very close to
$180\circ$: with this solid fraction a contact angle of
$179^\circ$ i sexpected, on the basis of the  Cassie relationship
\cite{callies05}, and assuming a contact angle of
$\theta_0=110^\circ$ on the solid posts. Contact angles $\sim
180^\circ$ are indeed reported by Choi and Kim for their
'nanoturf' surface \cite{choi06-b}. Note that the slip length is
strongly dependent on the solid fraction in this regime : for
example a slip length of $4.5\mu$m is predicted for a solid
fraction of $\phi_s=0.5\%$, corresponding to a contact angle of
$175^\circ$. More recently, and using also rheological
measurements, Truesdell {\it et al.} report huge slippage (in the
millimeter range) for surfaces with a strip geometry
\cite{truesdell06}, while $\phi_s\sim 50\%$ in their case. This
result is at odds with theoretical expectations and would require
further verifications. This points to the experimental
difficulties to measure slippage effects inherent to rheological
measurements, see Ref. \cite{bocquet06}. Eventually, we mention
the results of Joseph {\it et al.} who measured slip length on
superhydrophobic carbon nanotube carpets using micro-particle
velocimetry ($\mu$-PIV)measurements. Slip lengths in the micron
range are found and measured to be proportional to the underlying
pattern period $L$ (with fixed surface fraction
 of solid $\phi_s\sim 0.15$ in their case) \cite{joseph06}. This results is in
full agreement, both qualitative and quantitative, with
theoretical predictions of Eq. (\ref{slipplot}).

As an overall conclusion, these results and predictions do show
that very large slip lengths may be obtained only at the expense
of important efforts to obtain nano-engineered surfaces with very
small solid fraction. Typically, slip length larger than tens of
microns  are  expected only with micro-patterned surfaces for
which the measured contact angle is larger than $178^\circ$. This
is a very stringent condition.

\subsection{Effective boundary condition on porous surfaces}

In this context, it is interesting to point out the differences
and similarities between our  approach of ideal or structured
interfaces and the approach used in the field of porous media.
Effective slip  boundary conditions at the interface between a
bulk liquid and a porous media have been introduced empirically by
Beavers and Joseph \cite{joseph66,beavers67,goyeau03}, and
strongly resemble equation (\ref{bc1}). They can be justified on
the basis of matching a Brinkman description of the flow field in
the porous media
\begin{equation}\label{brinkman}
\eta_eff   \nabla^2 \mathbf{v} - \nabla P =  \frac{\eta}{K}
\mathbf{v}
\end{equation}
with the standard Stokes flow in the fluid. The resulting boundary
condition is
\begin{equation}\label{brinkman1}
\partial_z v_x (0) = \sqrt{\frac{\eta_{eff}/\eta}{K}} (v_x(0) - U)
\end{equation}
where  $K$ is the permeability of the porous medium, and $U=  -K
\nabla P$ is the drainage velocity deep inside the porous wall.
This condition is  often empirically modified to account for
variations in the local permeability or viscosity near the
interface \cite{goyeau03}. From a conceptual viewpoint, this
approach is different from the one described in section
\ref{ideal}, as it does not attribute physical properties to the
interface. Rather, the usual hydrodynamic approach is used
everywhere, with particular approximations relevant to porous
media. The approach described in the present section is somewhat
closer in spirit, with the difference that it uses essentially
exact results to describe the flow in the interfacial region.

\section{Analogy with thermal boundary conditions}
\label{thermal}


An interesting analogy can be established between the flow of a
fluid past a solid interface \cite{barrat03}, and the flow of heat
(or energy) transport across an interface. Let us  consider two
media ($1$ and $2$) separated by an ideal planar interface
perpendicular to the $z$ axis, and a steady heat flux ${\bf J}= J
{\bf e}_z$ perpendicular to the interface. Each medium is assumed
to be described by Fourier's law, with conductivity $\kappa_i$.
The temperature profile in medium $i$ is linear
\begin{equation}\label{tz}
    T_i(z) = \frac{J}{\kappa_i} z + T_i(0)
\end{equation}
The temperature jump at the interface, $T_1(0)-T_2(0)$, is the
exact analog of the slip velocity. Again, note that $T_i(0)$ can
be defined unambiguously as extrapolations of the bulk temperature
field in the interface region, without any reference to a local
temperature at the interface. A constitutive equation for the
interfacial region is written in analogy with equation (\ref{bc1})
\begin{equation}\label{rk1}
    T_i(0)-T_2(0)  = \frac{1}{R_K}{\bf J} \dot {\bf n}_{12}
\end{equation}
Equation (\ref{rk1}) defines the Kapitsa resistance $R_K$ (also
known as thermal boundary resistance), as the transport
coefficient describing the thermal effect of the interface. In
order to complete the analogy with the slip length, a "Kapitsa
length"  $\ell_K$ can be defined using a suitable thermal
conductivity, e.g. $\kappa_1$. $\ell_K= R_K\kappa_1$ is then the
the thickness of material $1$ which is thermally equivalent to the
interface.

This transport coefficient has been extensively studied in the
context of semiconductor physics, where the resistance can be
interpreted in terms of phonon scattering at the interface
\cite{swartz89}. It is also well known in the field of superfluid
liquid helium  \cite{swartz89} (where Kapitsa originally
introduced the notion), where the very large conductivity of
Helium II makes the effect particularly important. A Kubo formula
similar to (\ref{kubo}) can be formulated and reads
\cite{barrat03,puech86}
\begin{equation}\label{kubokapitsa}
    \frac{1}{R_K} =   \frac{S }{k_BT}\int_0^\infty dt \langle q(t)q(0)\rangle
\end{equation}
where $q(t)$ is the heat flux (per unit area)  across the
interface at time $t$, and $S$ is the interfacial area.

For simple fluid/solid interfaces, the effect of the Kapitsa
resistance has attracted much less attention. In the recent years
a number of numerical and experimental studies have appeared,
motivated in particular by the interest in thermal properties of
"nanofluids" (colloidal solution of oxide or metallic particles).
The general trend from simulation results \cite{barrat03,patel05}
is that the Kapitsa length will depend on the interfacial or
wetting properties in a way that is comparable to the slip length.
A weaker  affinity between the solid and the liquid, will result
in higher resistances. Other ingredients, that are irrelevant for
the slip length, include the difference in acoustic impedance
between the liquid ans the solid \cite{vladkov06}.
 Experimental studies involve
transient absorption or transient reflectivity experiments
\cite{wilson02,ge06}.
 Typical
values found   for $G=1/R_K$ are in the range
$50-300\mathrm{GW.m^{-2}.K^{-1}}$, with the smallest values
obtained for hydrophobic interfaces in contact with water
\cite{patel05,ge06}.  This corresponds to a Kapitsa length (using
the heat conductivity of water) of order $12\mathrm{nm}$, indeed
comparable to the slip length at similar interfaces. If an attempt
to model such a resistance through an "equivalent vapor layer" is
made, a thickness of the order of one molecular size is found.
This  shows  again that such transport coefficients are
intrinsically interfacial properties, that cannot be assigned to
the existence of a different "phase" at the interface.

We are not aware of any theoretical or experimental studies of
heat transport across structured interfaces similar to those
studied above for the slip properties. Obviously, strong effects,
involving thermally driven dewetting, could be expected in the
case of structured hydrophobic surfaces.

\section{the expected benefits of slippage}

\subsection{permeability and hydrodynamic dispersion}

One of the reason for the recent interest in slippage is that it
may facilitate flows in micrometric chanel, {\it e.g.}  for
microfluidic purposes. A simple calculation shows that for a
pressure drop flow in a slit with thickness $h$ and slip length
$b$ on the surfaces, the mean velocity is increased by a factor
$1+6 b/h$ as compared to the no-slip surfaces \footnote{A factor
$1+8b/h$ is found for a cylindrical channel}. Slippage thus
increases the permeability of channels and porous materials. A
second interesting property of slip flows is the reduction of
hydrodynamic dispersion of a probe transported in the channel
using a pressure drop flow. This dispersion originates in the
vanishing velocity at the surface, while it is maximum in the
middle of the channel. For slipping surfaces, the velocity does
not vanish at the surface, thereby reducing the difference between
the maximum and minimum velocities. A criterium for dispersion may
be defined as $\Delta v= (v_{max}-v_{min})/v_{max}$, which for a
slipping channel equals $\Delta v=(1+4b/h)^{-1}$.

For both phenomena, the efficiency of slippage is determined by
the ratio $b/h$. This shows that micrometric slip lengths have to
be reached for the slippage effects to be relevant for
microfluidic purposes. As discussed above, this requires a
specific engineering of the surfaces since slip lengths on bare
surfaces are well below the micron ($b\sim 20$nm for water on
hydrophobic surfaces). For example, for a channel with size $H\sim
10 \mu$m and a slip length $b\sim 20$nm, the increase of
permeability is of order of one percent ! For a slip length $b\sim
2\mu$m as measured {\it e.g.} in \cite{joseph06}, this
permeability is more than twice its value for no-slip surfaces.

\subsection{Interfacial transport phenomena}
\label{interfacial}



The above discussion relies on the fact that the flow is driven on
the scale of the channel $h$. However alternative methods do
generate flows within the interfacial structure at the
solid-liquid interface. Electro-osmosis, {\it i.e.} flow
generation by an electric field, is one such example which is
commonly used in microfluidics. As discussed in references
\cite{joly04,ajdari06,stone04} these interfacially driven
transport phenomena are considerably amplified by surface
slippage, {\it even for nanometric slip length}. The reason for
this amplification is that slip length $b$ now compares with the
thickness of the interfacial structure $\lambda$, and one expects
a amplification ratio for the transport of order $b/\lambda$. For
typical conditions, this ratio is larger than one even for
nanometric slip lengths and slippage strongly increase the
efficiency of  interfacial transport phenomena.

Let us precise the underlying mechanisms on the example of electro-osmosis.
Under the application of an electric field $E$, the solvent acquires a plug flow like velocity profile,
with a velocity $v_{EO}$
proportional to $E$ and given by the Smoluchowski formula~:
\begin{equation}
v_{EO}=-{\epsilon \zeta \over \eta} E
\label{veo}
\end{equation}
with $\epsilon$ the dielectric permittivity of the solvent, $\eta$
its viscosity; $\zeta$ is the Zeta potential of the surface, which
is traditionally assumed to match the electrostatic potential at
the position where the velocity profile vanishes (the shear-plane
position) \cite{hunter}.

\begin{figure}[t]
\centerline{\resizebox{7cm}{!}{\includegraphics{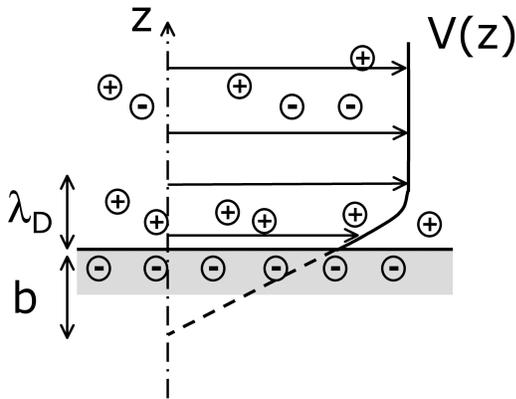}}}
\vspace*{-0.2cm} \caption[]{Schematic representation of
electro-osmotic flow past a slippery surface.} \label{fig3}
\end{figure}

Electro-osmosis originates within the Debye layer at the interface
\cite{hunter}. The Debye layer quantifies the width of the
interfacial regions at a charged surface and results from the
competition between the attraction of ions at the charged surface
and entropic effects. Its typical size $\lambda_D$ depends on the
salt concentration in the liquid, and is of the order of a few
nanometers : $\lambda_D\sim 3$nm for a concentration of salt of
$10^{-2}$ molar. The interfacial Debye layer is electrically
charged, while the remaining bulk solvent is electrically neutral.
Therefore, when applying an electric field, the electric driving
force is located within the Debye layer only. The driving force
per unit surface acting on the liquid thus writes $f_e=\sigma E$
with $\sigma$ the charge of the Debye layer, which is exactly
opposite to the surface charge $\sigma\simeq\epsilon
V_0/\lambda_D$, for weak surface potentials $V_0$. This leads to
$f_e\simeq -\epsilon {V_0\over \lambda_D} E$. Now the
electro-osmotic velocity results from the balance between this
electric driving force in the Debye layer and the viscous stress
at the surface. For a no-slip boundary condition, the latter is
$f_\eta\sim \eta v_{EO}/\lambda_D$, since the velocity varies on a
scale given by the Debye length. If slippage is exhibited at the
surface, the velocity varies on the size $\lambda_D+b$ and the
viscous stress becomes accordingly $f_\eta=\eta
v_{EO}/(\lambda_D+b)$, lower than for the no-slip BC. Gathering
these results we  get
\begin{equation}
\eta {v_{EO}\over \lambda_D+b} \sim - \epsilon {V_0\over \lambda_D} E
\end{equation}
This simple argument leads to the Smoluchowski formula, Eq.
(\ref{veo}), and provides an expression of the zeta potential in
the form
\begin{equation}
\zeta=V_0 \left(1 + {b\over \lambda_D}\right)
\label{zeta}
\end{equation}
This expression can be obtained using more rigourous description,
see \cite{stone04,joly04,joly06}. Molecular dynamics simulations
of charge transport confirm fully this relationship
\cite{joly04,joly06}. They show, furthermore, that for a
non-slipping surface $V_0$ can be identified with   the potential
at the plane of shear $V_0=V(z_s)$, with $z_s$ of the order of one
liquid layer. As expected a strong amplification of charge
transport is therefore demonstrated on slipping surfaces.

This behaviour can be generalized to all interfacial transport
phenomena. As practical examples, one may cite diffusio-osmosis
and thermo-osmosis. These two phenomena correspond to the
induction of a flow by the gradient of a solute concentration for
the former and by a gradient of temperature for the latter
\cite{anderson89}. The flow velocity for these phenomena is found
to be proportional to the applied gradient of the observable $O$
under consideration (electric potential, concentration,
temperature)~:
\begin{equation}
  v_s = - {1\over \eta}\Gamma L.[1+b/L].\frac{dO}{dx}
\end{equation}
$\Gamma$ and $L$  defined in terms of the (equilibrium) interfacial stress anisotropy,
$\Sigma_{eq}(x,y)=\sigma_n-\sigma_t$~:
 $\Gamma = \int_0^\infty \!\! dy .\frac{\partial\Sigma_{eq}}{\partial O}(y)$,
 and $L=\Gamma^{-1}\int_0^\infty\!\! dy .y.\frac{\partial \Sigma_{eq}}{\partial O}(y)$.
 These quantities are length scales of the order of the
range of interaction of the solute with the solid surface
\cite{ajdari06}

The origin of the phenomena thus lies in the thin interface layer of size $L$ where the solute interacts
specifically with the surface. For the diffusio-osmosis phenomena, the driving force is an osmotic
pressure gradient within the interface layer. For thermo-osmosis, this is surface energy gradient
induced by the temperature dependence of the stress anisotropy.

For all these phenomena, slippage again amplifies the transport phenomena with a ratio $b/L$.
Slippage, even nanometric, is therefore relevant to amplify interfacial transport in microfluidic
systems.



\section{Perspectives}

In this short review, we have tried to establish a clear picture
of our present understanding of slippage at different scales. At
the  molecular scale, we consider surfaces as ideal and introduce
an "intrinsic" constitutive relation for the interface. The
corresponding slip length depends strongly on the nature of
intermolecular interactions (hydrophobicity effect), but stays
within "molecular" orders of magnitude, a few ten nanometers at
most.

A different picture is obtained by considering surfaces structured
on an intermediate scale. To describe such situations, we adopt a
mesoscopic view of the interface, in which the molecular details
are hidden in a local, position dependent slip length. Such a view
can be validated through molecular  simulations. Hydrophobicity
effects are essential to reduce the contact area between solid and
liquid, so that a significant reduction of the solid-liquid
friction is obtained. When this reduction is achieved, much larger
slip lengths can be observed. The slip length is essentially
determined by the scale of the surface pattern and the surface
area of the "dewetted" part, on which a perfect slip boundary
condition applies at the local level.

This description at two different levels is necessary to describe
the full complexity of surfaces that can now be engineered at a
submicrometer  scale, with a combination of local hydrophobic
properties and of a complex geometric pattern. For patterns whose
typical scale is much larger than the intrinsic slip length, the
separation of lengths scales allows  a complete decoupling between
the molecular and the hydrodynamic descriptions. Molecular
simulations indicate that the same type of approach is possible,
even when the length scale of the pattern and the intrinsic slip
length are similar.

From this complete  theoretical description, it is clear that the
control of actual flow properties through surface engineering
necessitates a control of the surface state at each different
scale. When such a control is achieved, the flow of simple liquids
appear to be relatively well understood, both experimentally and
theoretically. More complex situations, such as the amplification
of interface driven flows   by molecular slippage described in
section \ref{interfacial}, have not been studied experimentally in
detail yet, but are promising  for the design of microfluidic
devices.

In section \ref{thermal} we discussed the analogy between
"temperature slip" and "velocity slip". Experiments and
simulations have already shown that this analogy can be made
quantitative, at least when ideal surfaces are concerned. The role
of nano-structuration of the surfaces for thermal transport has
not been explored up to now. Nonlinear effects may be expected for
strong heat fluxes, while cross effects between heat and momentum
transport are expected to be negligible, based on symmetry
arguments.

Finally, it could be interesting, at least from a formal point of
view, to extend the above discussion of heat and momentum
transport across an interface to diffusive mass transport. The
standard description assumes a local partition equilibrium at the
interface, or equivalently a continuity of the chemical potential.
Following the approach outlined here, it would seem more
consistent to introduce a discontinuity in chemical potential
proportional to the diffusive flux across the interface. The
magnitude of this discontinuity has not, to our knowledge, be
explored at the molecular scale.

\begin{acknowledgments}
We thank A. Ajdari, C. Barentin, E. Charlaix, C. Cottin-Bizonne,
P. Joseph, L. L\'eger, and C. Ybert
 for many stimulating discussions.

\end{acknowledgments}

\bibliography{biblio1}

\end{document}